\documentclass[doublecol]{epl2} 
\usepackage{amssymb}
\usepackage{amsmath}
\usepackage{amsfonts}

\title{Exact relationship between the entanglement entropies of XY and  quantum Ising chains}
\shorttitle{Exact relationship}
\author{Ferenc Igl\'oi\inst{1,2} \and R\'obert Juh\'asz\inst{1}}
\shortauthor{F. Igl\'oi \etal}
\institute{
\inst{1} Research Institute for Solid
State Physics and Optics, H-1525 Budapest, P.O.Box 49, Hungary \\
\inst{2}
Institute of Theoretical Physics,
Szeged University, H-6720 Szeged, Hungary
}


\abstract{
We consider two prototypical quantum models, the spin-1/2 XY chain and
the quantum Ising chain and study their entanglement entropy, $S(\ell,L)$, of blocks of $\ell$ spins
in homogeneous or inhomogeneous systems of length $L$. By using two different approaches,
free-fermion techniques and perturbational expansion,
an exact relationship between the entropies is revealed. 
Using this relation we translate known results between
the two models and obtain, among others, the additive constant of the
entropy of the critical homogeneous quantum Ising chain and the effective central charge of the random XY chain.
}

\begin{document}
\maketitle

\newcommand{\bc}{\begin{center}}
\newcommand{\ec}{\end{center}}
\newcommand{\be}{\begin{equation}}
\newcommand{\ee}{\end{equation}}
\newcommand{\beqn}{\begin{eqnarray}}
\newcommand{\eeqn}{\end{eqnarray}}


\vskip 2cm

\section{Introduction}

Recently, we have witnessed a growing interest in entanglement effects in
quantum many-body systems\cite{fazio}. If an isolated quantum system is divided into two parts, $\cal A$ and $\cal B$, all information about $\cal A$ is contained in the
reduced density matrix: $\rho_{\cal A}=Tr_{\cal B} |0\rangle \langle 0
|$, where $|0\rangle$ denotes the ground state of the system. 
To quantify the quantum entanglement between
$\cal A$ and $\cal B$, different measures have been introduced; a
frequently used quantity is the von Neumann entropy defined by: 
\be
S_{\cal A}=-\textrm{Tr} (\rho_{\cal A} \log_2 \rho_{\cal A}) = -\textrm{Tr} (\rho_{\cal B} \log_2 \rho_{\cal B}).
\ee
Contrary to the thermal entropy, the entanglement entropy is not an extensive quantity, but
for a non-critical system it scales with the area of the interface\cite{AREA} separating the block
(i.e. $\cal A$) and the environment ($\cal B$). 
In the case of a one-dimensional (1d) infinite system  
and a finite block of length, $\ell$, 
the interface consists of a few points and the
entanglement entropy generally approaches a finite value  
in the limit $\ell\to\infty$. 
At the critical point, however, $S_{\cal A}$ diverges logarithmically
as   
\be
S_{\cal A}= \frac{c}{3} \log_2 \ell + c_1,
\label{S_l}
\ee
where the prefactor, $c$, is universal in homogeneous systems and given by the central charge of the associated conformal field theory\cite{calabrese_cardy}, whereas the constant, $c_1$, is non-universal. 
The relation in Eq.(\ref{S_l}) has been generalized by Calabrese and Cardy\cite{calabrese_cardy}
for finite systems,
for different boundary conditions, for non-critical
systems in the vicinity of the transition point and for finite temperatures. In higher dimensions
only a few results are available, mainly for non-interacting fermions\cite{FREE-F} and bosons\cite{BOSON-NUM,BOSON-ANA} but also
random quantum Ising models have been investigated\cite{lri07}.

Some of the conformal results have been tested on integrable quantum spin chains in
particular on the antiferromagnetic $XY$-chain, defined by the
Hamiltonian\cite{lsm}:
\be
{\cal H}_{XY} =
\sum_{i=1}^L (J_{i}^x S_i^x S_{i+1}^x+J_{i}^y S_i^y S_{i+1}^y)\;.
\label{eq:H_XY}
\ee
Here the $S_i^{x,y}$'s are spin-1/2 operators at site $i$, $S_{L+1}^{x,y} \equiv S_1^{x,y}$ and 
the couplings $J_{i}^x$ and $J_{i}^y$ may be different and 
site-dependent in general.   
In the followings, we restrict ourselves to even system sizes $L$. 
If the interaction is isotropic on average in the sense that 
$[\ln J^x]_{av}=[\ln J^y]_{av}$ holds, where $[\dots ]_{av}$
stands for the average over the distribution of couplings, 
the model is critical, which manifests itself in the vanishing of the
gap and the algebraic decay of correlation functions
in the thermodynamic limit $L \to \infty$.
For the special case of the homogeneous $XX$-chain, i.e. for 
$J_i^x=J_{i}^y=J$ both the prefactor
($c=1$) and the constant $c_1$ in Eq.(\ref{S_l}) 
have been calculated exactly\cite{jim_korepin}.

Another basic one-dimensional quantum model is the quantum
Ising chain (QIC) defined by the Hamiltonian
\be
{\cal H}_I =
-\frac{1}{2}\sum_{i=1}^L \lambda_{i}\sigma_i^x \sigma_{i+1}^x-\frac{1}{2}\sum_{i=1}^L h_i \sigma_i^z
\label{eq:H_I}
\ee
in terms of the Pauli-operators $\sigma_i^{x,z}=2S_i^{x,z}$ at site $i$ ($\sigma_{L+1}^{x,z} \equiv \sigma_1^{x,z}$)
and $\lambda_i$ and $h_i$ are nearest neighbor couplings and the
transverse fields, respectively.
This model exhibits a quantum phase transition if 
$[\ln \lambda]_{av}=[\ln h]_{av}$, i.e. for the homogeneous system
with $\lambda_i=\lambda$ and $h_i=h$, the critical point is
located at $J=h$ \cite{pfeuty}.
The saturation value of the von Neumann entropy in the homogeneous model
has been found to exhibit
a singularity in the vicinity of the
critical point as $S_{\infty}=\frac{c}{3} \log_2 \xi$, where $\xi$
denotes the correlation length diverging at criticality
\cite{calabrese_cardy}. The prefactor is obtained to be $c=1/2$ in an 
analytical calculation using a mapping between the reduced
density matrix of the model and the corner transfer matrix of the 2d classical Ising model\cite{calabrese_cardy,peschel04}.

The entanglement entropies of the XX and the quantum Ising model 
have also been studied in the presence of
quenched disorder, when the parameters of the models 
(the isotropic couplings $J^x_i=J^y_i\equiv J_i$,
or the $\lambda_i$ bonds and the $h_i$ transverse fields, respectively)
are independent and identically distributed
random numbers. The average entropy has been calculated analytically\cite{refael} by a strong disorder renormalization
group method\cite{review}, which is found to follow the logarithmic law in Eq.(\ref{S_l}). In this case
the prefactor, which is called the effective central charge and
denoted by $c_{\rm eff}$ is
obtained as $c_{\rm eff}(XX)=\ln 2$ for the random $XX$-chain and $c_{\rm eff}(I)=\ln 2/2$ for the random critical QIC,
respectively.

Numerical studies of the entanglement entropy were performed on different XX-chains,
in which the effect of a free boundary\cite{boundary}, a defect coupling\cite{XX_defect},
random\cite{laflorancie} or aperiodic interactions\cite{ijz07}
and the presence of an energy current in the system\cite{ez05},
etc. were investigated. 
For the random QIC, the location of the maxima
of the entropy is used to define sample-dependent critical points\cite{ilrm07}.
The evolution of the entropy after a quench in both models is also the subject of recent
investigations\cite{cc_quench,i_quench}.

It is known for some time that the two Hamiltonians in Eqs.(\ref{eq:H_XY}) and (\ref{eq:H_I})
can be mapped into each other through a canonical transformation\cite{peschel_schotte,fisherxx,ijr00},
which is described in the Appendix.
As a consequence, the spectrum of the two Hamiltonians, as well as some correlation functions
of the two models are related, as well. 
One might ask the question, whether a similar correspondence
can be found concerning the entanglement entropies of the two models. 
At first thought, the existence of
such a relation is not obvious since the transformation of the operators
in Eqs.(\ref{mapping}) and (\ref{mapping_inv}) is nonlocal, 
c.f. operators in ${\cal A}$
for one model are expressed with operators located both in ${\cal A}$ and ${\cal B}$.

In this paper, we study in detail the relation between the entanglement entropies of the
two models by two approaches. In the first approach
we calculate the entanglement entropies in the free-fermion representation\cite{lsm}.
By this method, the Hamiltonians are first expressed in terms of free fermions,
which requires the solution of an eigenvalue problem of dimensions, $L \times L$,
then after a second transformation, the systems assume the form of non-correlated fermionic modes,
which is obtained by solving an $\ell \times \ell$ eigenvalue problem.
These eigenvalue problems are then compared for the two models.
In the second approach, the entanglement entropy is calculated 
by a perturbation expansion, in terms of
different powers of the coupling term connecting the two parts of the
system and the expressions obtained for the two
models are then compared. 
The relation between the entropies is then used to transfer
existing results between the two models.

\section{Free-fermion calculation}

The main steps of the calculations to be carried out in this section
are summarized as follows. 
The Hamiltonians are expressed in terms of fermion operators and 
are diagonalized. Then, the restricted correlation matrix is
constructed and it is transformed to a form corresponding to 
non-correlated fermionic modes by a canonical transformation. 
Finally, the entanglement entropy is calculated from the eigenvalues
of this matrix.

\subsection{Diagonalization of the Hamiltonians}

Both models can be expressed in terms of fermion creation and
annihilation operators $c^+_i$ and $c_i$, respectively \cite{lsm}, 
which are obtained through the Jordan-Wigner transformation:
$a_j^{\pm}=S_j^x \pm iS_j^y$ and $c^+_i=a_i^+\exp\left[\pi i \sum_{j}^{i-1}a_j^+a_j^-\right]$,
$c_i=\exp\left[\pi i \sum_{j}^{i-1}a_j^+a_j^-\right]a_i^-$. The Hamiltonian of the XY-chain
in Eq.(\ref{eq:H_XY}) is expressed as:
\beqn
{\cal H}_{XY}
 &=&
\sum_{i=1}^{L-1} \frac{1}{4} \left\{(J_i^x-J_i^y)c^+_i c^+_{i+1}+(J_i^x+J_i^y)c^+_i c_{i+1} + {\rm h.c.}
\right\}\cr&-&\frac{1}{4} w\left\{(J_L^x-J_L^y)c^+_L c^+_{1}+(J_L^x+J_L^y)c^+_L c_{1} + {\rm h.c.} \right\}
\label{ferm_XY}
\eeqn
where $w=\exp(i\pi N_c)$, with $N_c=\sum_{i=1}^L c^+_i c_i$. 

Similarly one obtains for the Hamiltonian of the QIC\cite{pfeuty} in Eq.(\ref{eq:H_I}):
\beqn
{\cal H}_{I}&=&
\sum_{i=1}^{L}h_i\left( c^+_i c_i-\frac{1}{2} \right) -
\frac{1}{2}\sum_{i=1}^{L-1}\lambda_i(c^+_i-c_i)(c^+_{i+1}+c_{i+1})\cr
&+&\frac{1}{2}w\lambda_L(c^+_L-c_L)(c^+_{1}+c_{1}).
\label{ferm_I}
\eeqn
For both Hamiltonians, 
the $N_c$ is even in the ground state, thus we have $w=1$.
Moreover, both Hamiltonians 
are quadratic in $c^+_i$ and
$c_i$, therefore they can be diagonalized by standard techniques\cite{lsm}.
In our approach, 
the basic quantity is a $2L \times 2L$ matrix, denoted by $\mathbf{T}$,
which can be interpreted as the transfer matrix
of directed walks and which has been introduced in\cite{it96} for the QIC and in\cite{ijr00}
for the XY-chain\cite{note1}.
From the solution of the eigenvalue problem of $\mathbf{T}$ one obtains both the
energies of the free-fermionic modes, $\Lambda_k>0$, $k=1,2,\dots L$, and two sets of vectors:
${\mathbf \Phi}_k$ and ${\mathbf \Psi}_k$, both having $L$ components\cite{note}.
For the XY-chain, $\mathbf{T}$
is given by\cite{ijr00}:
\begin{widetext}
\be
\mathbf{T}_{XY}=
\begin{pmatrix}
 0 & 0 & J_1^y &      &       &    &&    & -wJ_L^x & 0      \cr
0      & 0     &   0   & J_1^x&       &    &&    &         &-wJ_L^y \cr
 J_1^y&  0      &    0   &  0   & J_2^x &    &&    &         &                   \cr
       & J_1^x  & 0    &   0   & 0     & J_2^y  &  &&       &                    \cr
       &        & J_2^x &  0   & 0     &  0     &  J_3^y & &&                \cr
       &        &     &  J_2^y   & 0     &  0     & 0 &\ddots  & &                \cr
       &        &     &  &\ddots &\ddots&\ddots   & &   J_{L-1}^y&              \cr
       &        &      &     &  &  J_{L-2}^y &  0 & 0 & 0 &  J_{L-1}^x   \cr
 -wJ_L^x&        &    &&&     & J_{L-1}^y   & 0  & 0     & 0     \cr
0     & -wJ_L^y &       &   &    &&      & J_{L-1}^x &  0   & 0          \cr
\label{TXY}
\end{pmatrix}
\ee
\end{widetext}
\begin{floatequation}
\mbox{\textit{see eq.~\eqref{TXY} above}}
\nonumber
\end{floatequation}
and the eigenvectors contain the components: $(\Phi_k(1),\Psi_k(1),\Psi_k(2),\Phi_k(2),\Phi_k(3),\Psi_k(3)\dots,\Phi_k(L-1)$, $\Psi_k(L-1),\Psi_k(L),\Phi_k(L))$.

For the QIC the $\mathbf{T}_{I}$ matrix reads as\cite{it96}:
\be
\mathbf{T}_{I}=
\begin{pmatrix}
0      &  h_1 &      &       &        & -w\lambda_L       \cr
 h_1 & 0      &\lambda_1 &       &     &                   \cr
       & \lambda_1  & 0    & h_2&     &                    \cr
       &        &   \ddots  &\ddots &\ddots   &                \cr
       &        &         & \lambda_{L-1}   & 0       &   h_L    \cr
-w\lambda_L &       &   &          & h_L &  0             \cr
\end{pmatrix}
\label{TQIC}
\ee
and the eigenvectors have the components: $(-\Phi_k(1),\Psi_k(1),-\Phi_k(2),\Psi_k(2),\dots,-\Phi_k(L),\Psi_k(L))$.

As shown in\cite{ijr00} the $L$ eigenvectors of $\mathbf{T}_{XY}$ with
positive eigenvalues can be divided into two classes:

i) For the first class of vectors, which we mark with odd superscripts
we have $\Phi_{2k-1}(2i)=\Psi_{2k-1}(2i-1)=0$, whereas the
non-zero components of the vectors are obtained from the eigenvalue problem of the matrix:
\be
\mathbf{T^{(\sigma)}}_{I}=
\begin{pmatrix}
0      &  J_1^y &      &       &        & -wJ_L^x       \cr
J_1^y  & 0      &J_2^x &       &     &                   \cr
       & J_2^x  & 0    & J_3^y &     &                    \cr
       &        &   \ddots  &\ddots &\ddots   &                \cr
       &        &         & J_{L-2}^x   & 0       &   J_{L-1}^y    \cr
-wJ_L^x &       &   &          & J_{L-1}^y &  0             \cr
\end{pmatrix}
\label{Tsigma}
\ee
and the eigenvectors have the components: $(\Phi_k(1),\Psi_k(2),\Phi_k(3),\Psi_k(4),\dots,\Phi_k(L-1),\Psi_k(L))$.

This is just the $\mathbf{T}$ matrix of a QIC with the Hamiltonian
$H_I(\sigma)$ given in Eq.(\ref{mapping_H}).
Denoting the components of the vectors corresponding to this
QIC by $\Phi_{k}^{(\sigma)}(i)$ and $\Psi_{k}^{(\sigma)}(i)$ we have the correspondences:
\be
\Phi_{2k-1}(2i-1)=-\Phi_{k}^{(\sigma)}(i),\quad \Psi_{2k-1}(2i)=\Psi_{k}^{(\sigma)}(i)\;.
\label{rel:2k-1}
\ee
ii) For the eigenvectors of the second class, which are marked with even
superscripts, the vanishing components are $\Phi_{2k}(2i-1)=\Psi_{2k}(2i)=0$, whereas the
non-zero components of the vectors are obtained from the eigenvalue problem of the matrix:
\be
\mathbf{T^{(\tau)}}_{I}=
\begin{pmatrix}
0      &  J_1^x &      &       &        & -wJ_L^y       \cr
J_1^x  & 0      &J_2^y &       &     &                   \cr
       & J_2^y  & 0    & J_3^x &     &                    \cr
       &        &   \ddots  &\ddots &\ddots   &                \cr
       &        &         & J_{L-2}^y   & 0       &   J_{L-1}^x    \cr
-wJ_L^y &       &   &          & J_{L-1}^x &  0             \cr
\end{pmatrix}
\ee
and the eigenvectors have the components: $(\Psi_k(1),\Phi_k(2),\Psi_k(3),\Phi_k(4),\dots,\Psi_k(L-1),\Phi_k(L))$.

This is again the $\mathbf{T}$ matrix of a QIC with the Hamiltonian
$H_I(\tau)$ given in Eq.(\ref{mapping_H}).
Denoting the components of the vectors corresponding to this
QIC by $\Phi_{k}^{(\tau)}(i)$ and $\Psi_{k}^{(\tau)}(i)$ we have the correspondences:
\be
\Phi_{2k}(2i)=\Psi_{k}^{(\tau)}(i),\quad \Psi_{2k}(2i-1)=-\Phi_{k}^{(\tau)}(i)\;.
\label{rel:2k}
\ee
Thus we conclude that the explicit solution of $\mathbf{T}_{XY}$
requires the diagonalization
of two $\mathbf{T}$-s of QIC-s, which is just equivalent to the mapping
described in the Appendix.

\subsection{Correlation matrix and entanglement entropy}

Next we consider a block of length $\ell$, consisting of spins $i=1,2,\dots,\ell$ and the
reduced density matrix is given by: ${\mathbf \rho}_{\ell}=Tr_{L-\ell} |0\rangle \langle 0 |$. 
For free fermionic systems ${\mathbf \rho}_{\ell}$ can be reconstructed from the
restricted correlation matrix\cite{peschel,vidal}, ${\mathbf G}$, the matrix-elements of which are given by:
\beqn
G_{m,n}=\langle 0 |(c^+_n-c_n)(c^+_m+c_m)|0 \rangle\cr
=-\sum_{k=1}^L \Psi_k(m) \Phi_k(n),\quad m,n=1,2,\dots,\ell
\label{G}
\eeqn
For the XY chain using the properties of the
vectors, ${\mathbf \Phi}_k$ and ${\mathbf \Psi}_k$ in Eqs.(\ref{rel:2k-1}) and (\ref{rel:2k})
we obtain for the matrix-elements:
\beqn
G_{2i,2j}&=&0,\quad\quad G_{2i-1,2j-1}=0 \cr
G_{2i,2j-1}&=&-G_{i,j}^{(\sigma)},\quad G_{2i-1,2j}=-G_{j,i}^{(\tau)}\;,
\label{G_XY}
\eeqn
where $G_{i,j}^{(\sigma,\tau)}$ denotes the matrix-element of the correlation matrix
of the QIC with Hamiltonian $H_I(\sigma,\tau)$.
The correlation matrix for even $\ell$ 
is bipartite, being composed of $2 \times 2$ matrices
\be
\left[\begin{matrix}
0 & -G^{(\tau)}_{j,i}                       \cr
-G^{(\sigma)}_{i,j}        &        0         \cr
\end{matrix}\right]\;,
\label{bipartite}
\ee
$i,j=1,2,\dots \ell/2$.

In order to obtain the von Neumann entropy, $S(\ell,L)$, one diagonalizes ${\mathbf \rho}_{\ell}$,
which is given through a canonical transformation:
\be
\mu_q=\sum_{i=1}^l\left[ \frac{1}{2}\left(v_q(i)+u_q(i)\right)c_i+\frac{1}{2}\left(v_q(i)-u_q(i)\right)c_i^+\right]
\ee
where the $v_q(i)$ and $u_q(i)$ are real and normalized: $\sum_i^{\ell} v_q^2(i)=\sum_i^{\ell} u_q^2(i)=1$.
In the transformed basis we have
\be
\langle 0 |\mu_q \mu_p|0 \rangle=0,\quad \langle 0 |\mu_q^+ \mu_p|0 \rangle=\delta_{qp}\frac{1+\nu_q}{2}\;,
\ee
for $p,q=1,2,\dots \ell$. Thus the fermionic modes are uncorrelated and the eigenvalues of
${\mathbf \rho}_{\ell}$
are the products of $(1 \pm \nu_q)/2$, $q=1,2,\dots \ell$. The entropy of the system is given by
the sum of binary entropies:
\be
S(\ell,L)=-\sum_{q=1}^{\ell} \left[\frac{1+\nu_q}{2} \log_2 \frac{1+\nu_q}{2}
+\frac{1-\nu_q}{2} \log_2 \frac{1-\nu_q}{2}\right]
\label{binary}
\ee
The $\nu_q$-s are the solution of the equations:
\be
{\mathbf G}{\mathbf u}_q=\nu_q {\mathbf v}_q,\quad
{\mathbf G}^T{\mathbf v}_q=\nu_q {\mathbf u}_q\;,
\ee
or, equivalently, one has
\be
{\mathbf G}{\mathbf G}^T{\mathbf v}_q=\nu_q^2 {\mathbf v}_q,\quad
{\mathbf G}^T{\mathbf G}{\mathbf u}_q=\nu_q^2 {\mathbf u}_q\;.
\label{G_GT}
\ee
For the XY chain with even $\ell$, the matrix ${\mathbf G}{\mathbf G}^T$  
is composed of $2 \times 2$ diagonal matrices
\be
\left[\begin{matrix}
[(\mathbf{G}^{(\sigma)})^T \mathbf{G}^{(\sigma)}] _{i,j}                & 0         \cr
0             &        [\mathbf{G}^{(\tau)} (\mathbf{G}^{(\tau)})^T] _{i,j}         \cr
\end{matrix}\right]\;,
\label{G_even}
\ee
thus ${\mathbf G}{\mathbf G}^T$ can be written in a form which consists of two diagonal blocks 
and the eigenvalues
are obtained by solving two separate eigenvalue problems of
$(\mathbf{G}^{(\sigma)})^T \mathbf{G}^{(\sigma)}$ and $(\mathbf{G}^{(\tau)})^T \mathbf{G}^{(\tau)}$. 
Now, it follows from Eq.(\ref{binary}) that the entanglement
entropy of the XY-chain in Eq.(\ref{eq:H_XY}) is the sum of the entanglement entropies of the two
QIC-s defined in Eq.(\ref{mapping_H}):
\be
S^{(XY)}(\ell,L)=S^{(\sigma)}(\ell/2,L/2)+S^{(\tau)}(\ell/2,L/2)\;.
\label{S_rel}
\ee
This relation constitutes the main result of our paper.

One can make sure easily that the above result holds not only for
blocks of contiguous spins but for any block composed of pairs 
 of adjacent sites $(2i-1,2i)$.
In any other case, the matrix ${\mathbf G}{\mathbf G}^T$ 
is still block-diagonal and the von Neumann entropy can be written as
a sum of two terms each of which depend exclusively on the parameters
of one of the decoupled QIC's, however, these terms are no longer
to be interpreted as entropies of some blocks in the QIC's.  

Next, we turn to the case of odd $\ell$,
when the correlation matrix is obtained by leaving out the last row
and column of the
matrix with $\ell+1$, thus the structure in Eq.(\ref{bipartite}) is lost. 
The matrix ${\mathbf G}$ consists of non-square submatrices of size
$l\times (l+1)$ and $(l+1)\times l$ of $\mathbf{G}^{(\tau)}$ and
$\mathbf{G}^{(\sigma)}$, respectively.  
Consequently, the eigenvalues of ${\mathbf G}{\mathbf G}^T$ can not be expressed
by those obtained from the two QIC-s. Then the relation in Eq.(\ref{S_rel}) is
only asymptotically valid, as $\ell \gg 1$ and the corrections are
of the order of $1/\ell$.

\section{Perturbative calculation}

In this section we split the Hamiltonian of the system as:
\be
{\cal H}={\cal H}_{\cal A}+{\cal H}_{\cal B}+{\cal V}\;,
\label{H_split}
\ee
where ${\cal H}_{\cal A}$ and ${\cal H}_{\cal B}$ are the Hamiltonians of
the free subsystems, ${\cal A}$ and ${\cal B}$, respectively, and ${\cal V}$
is the interaction term, which reads for the XY-chain as:
\beqn
{\cal V}_{XY}&=&H(\ell)+H(L) \cr
H(\ell)&=&J_{\ell}^x S_{\ell}^x S_{\ell+1}^x+J_{\ell}^y S_{\ell}^y S_{\ell+1}^y \cr
H(L)&=&J_{L}^x S_{L}^x S_{1}^x+J_{L}^y S_{L}^y S_{1}^y\;.
\label{V_XY}
\eeqn
Let us denote the eigenstates of ${\cal H}_{\cal A}$ by $|\varphi_{i}^{\cal A}\rangle$
with energies $E_i^{\cal A}$, and similarly for ${\cal H}_{\cal B}$
the eigenstates are $|\varphi_{k}^{\cal B}\rangle$ with eigenvalues $E_k^{\cal B}$.
The ground state of the total system with Hamiltonian ${\cal H}$ can be expressed in terms of
$|\varphi_{i}^{\cal A}\rangle$ and $|\varphi_{k}^{\cal B}\rangle$ as 
$|0\rangle=\sum_i^{\cal A} \sum_k^{\cal B} c(i,k)(|\varphi_{i}^{\cal A}\rangle \otimes |\varphi_{k}^{\cal B}\rangle)$,
so that the reduced density matrix, ${\mathbf \rho}_{\ell}$, has the matrix-elements:
\beqn
\langle \varphi_{i}^{\cal A}|{\mathbf \rho}_{\ell}|\varphi_{j}^{\cal A}\rangle={\mathbf \rho}_{\ell}(i,j)&=&
\sum_k^{\cal B}(\langle \varphi_{i}^{\cal A}|\otimes\langle \varphi_{k}^{\cal B}|)0\rangle\langle 0(|\varphi_{k}^{\cal B}\rangle\otimes|\varphi_{j}^{\cal A}\rangle)\cr
&=&\sum_k^{\cal B} c(i,k)c^*(j,k)\;.
\eeqn
Here $c(0,0)=1$, otherwise
the expansion coefficients $c(i,k)$ are calculated perturbatively.
In leading order we have
\be
c(i,k)=-\frac{(\langle \varphi_{i}^{\cal A}|\otimes\langle \varphi_{k}^{\cal B}|){\cal V}(|\varphi_{0}^{\cal B} \rangle \otimes |\varphi_{0}^{\cal A}\rangle)}
{E_i^{\cal A}+E_k^{\cal B}-E_0^{\cal A}-E_0^{\cal B}} +\dots\;
\ee
and the higher order terms are sums of products containing factors of the form:
\be
f(i,i'|k,k')=
\frac{(\langle \varphi_{i}^{\cal A}|\otimes\langle \varphi_{k}^{\cal B}|){\cal V}(|\varphi_{k'}^{\cal B}\rangle \otimes|\varphi_{i'}^{\cal A}\rangle)}
{E_i^{\cal A}+E_k^{\cal B}-E_{i'}^{\cal A}-E_{k'}^{\cal B}}\;,
\label{factor}
\ee
such that $E_i^{\cal A}+E_k^{\cal B}-E_{i'}^{\cal A}-E_{k'}^{\cal B} \ne 0$.
For the XY-chain with the interaction term in Eq.(\ref{V_XY}) we have $f(i,i'|k,k')=f_{\ell}(i,i'|k,k')+f_L(i,i'|k,k')$, in which the first term is given by:
\beqn
f_{\ell}(i,i'|k,k')&=&\frac{1}{\Delta E}[J_{\ell}^x \langle\varphi_{i}^{\cal A}|S_{\ell}^x|\varphi_{i'}^{\cal A}\rangle
 \langle\varphi_{k}^{\cal B}|S_{\ell+1}^x|\varphi_{k'}^{\cal B}\rangle \cr
&+&J_{\ell}^y \langle\varphi_{i}^{\cal A}|S_{\ell}^y|\varphi_{i'}^{\cal A}\rangle \langle\varphi_{k}^{\cal B}|S_{\ell+1}^y|\varphi_{k'}^{\cal B}\rangle]
\label{f_1}
\eeqn
with $\Delta E=E_i^{\cal A}+E_k^{\cal B}-E_{i'}^{\cal A}-E_{k'}^{\cal B}$ and there is a similar
expression for $f_L(i,i'|k,k')$, as well. Note that the matrix-elements in these expressions are separated
as the product of two matrix-elements of the surface operators in the two subsystems.

In the next step we perform the same perturbation expansion in terms of two independent
QIC-s with parameters given in Eq.(\ref{mapping_H}), 
in which case for even $\ell$ the
perturbation is located at $\ell/2$ and $L/2$ and given by:
\beqn
\tilde{H}(\ell/2)&=&\frac{1}{4}J_{\ell}^x \sigma_{\ell/2}^x \sigma_{\ell/2+1}^x+\frac{1}{4}J_{\ell}^y \tau_{\ell/2}^x \tau_{\ell/2+1}^x \cr
\tilde{H}(L/2)&=&\frac{1}{4}J_{L}^x \sigma_{L/2}^x \sigma_{1}^x+\frac{1}{4}J_{L}^y \tau_{L/2}^x \tau_{1}^x \;.
\label{V_I}
\eeqn
Using the mapping in Eq.(\ref{mapping_inv}) we obtain $H(\ell)=\tilde{H}(\ell/2)$
and
\be
H(L)=\frac{1}{4}J_{L}^x \sigma_{L/2}^x \sigma_{1}^x\prod_{j=1}^{L/2}\tau_j^z+\frac{1}{4}J_{L}^y \tau_{L/2}^x \tau_{1}^x \prod_{j=1}^{L/2}\sigma_j^z\;.
\ee
For the QIC calculation we denote by $\tilde{f}_{\ell/2}(i,i'|k,k')$ the factor, in which we
use the states labeled by $i,i'|k,k'$ in Eq.(\ref{factor}). Then $\Delta E$ remains the same, as well as
the matrix-elements are invariant, so that ${f}_{\ell}(i,i'|k,k')=\tilde{f}_{\ell/2}(i,i'|k,k')$.
For the other term, $\tilde{f}_{L/2}(i,i'|k,k')$, the only difference, that the transformed
perturbation contains also the products
$\prod_{j=1}^{L/2}\tau_j^z$ and $\prod_{j=1}^{L/2}\sigma_j^z$, which commute with the Hamiltonians
$H_I(\tau)$ and $H_I(\sigma)$, respectively, and have the eigenvalues $p=\pm 1$. The excited states, however, which
enter into the expansion of $c(i,k)$ have the same parity as the ground state, $p=1$, thus also ${f}_{L}(i,i'|k,k')=\tilde{f}_{L/2}(i,i'|k,k')$.

We conclude that for even $\ell$ the expansion coefficients $c(i,k)$ are
identical for the XY model as well as for two independent QIC-s with parameters given in 
Eq. (\ref{mapping_H}) in all order of the perturbation expansion. 
Consequently, the same is true for the elements of the reduced
density matrix and finally for the entanglement entropy. 
In this way, we have reobtained the
result already calculated in Eq.(\ref{S_rel}).

If the size of the cell, $\ell$ is odd, then the interaction term in Eq.(\ref{V_XY})
is transformed as:
\be
H(\ell)=\frac{1}{4}J_{\ell}^x \sigma_{(\ell+1)/2}^z+\frac{1}{4}J_{\ell}^y \tau_{(\ell+1)/2}^z 
\ee
and similarly for $H(L)$. These cannot be written in a
separated form in terms of the $\sigma$ and $\tau$ operators, 
thus Eq.(\ref{S_rel}) is no longer valid.

\section{Discussion}

In this paper we have derived an exact relation in Eq.(\ref{S_rel}) between the entanglement entropy
of the XY-chain and that of the QIC. This relation is valid for a finite block of even size
and holds also for inhomogeneous couplings. Since the derivation is based on a mapping between the
two correlation matrices similar relations can be obtained for another measures of the entanglement,
such as the R\'enyi entropy or the concurrence.

Before discussing the simple consequences of the relation in
Eq.(\ref{S_rel}) we begin with the comb entanglement\cite{comb} of the XY-chain, when
the block $\cal A$ consists of $\ell \le L/2$ spins,
which occupy sites having the same parity. Then, according to Eq.(\ref{G_XY})
all elements of the matrix $\mathbf{G}$ are zero and the entropy is $S(\ell,L)=\ell$,
so that the block is
maximally entangled with the environment. 
In case of strongly disordered $XX$ chains, which possess
asymptotically a random singlet ground state\cite{review}, 
this finding can be obtained directly since singlet bonds form
exclusively between spins at sites with different parities. 
As can be seen, the singlet-state approximation happens to give an
exact result for the comb entanglement.

First we consider a homogeneous $XY$ chain with $J_x=(1+\gamma)$ and
$J_y=(1+\gamma)$, so that in the equivalent decoupled QIC-s we have $\lambda=1$
and $h=(1-\gamma)/(1+\gamma)$ ($h=(1+\gamma)/(1-\gamma)$) for the $\sigma$ ($\tau$) chain.
In the thermodynamic limit according to Eq.(\ref{S_rel}) we have:
\be
S^{(XY)}(\gamma)=
S^{(I)}\left(h=\frac{1-\gamma}{1+\gamma}\right)+S^{(I)}\left(h=\frac{1+\gamma}{1-\gamma}\right)\;,
\ee
which is indeed satisfied with the known exact results\cite{jim_korepin,peschel04}. For
the XX-chain, which corresponds to $\gamma=0$, the entropy is just the double of the
entropy of the critical QIC. Consequently, the
central charges are related as $c(XX)=2 c(I)$, which is indeed observed in the
exact calculations\cite{jim_korepin,calabrese_cardy,peschel04}.
For the non-universal constant in Eq.(\ref{S_l}) we have the relation $c_1(XX)=2 [-\frac{c(I)}{3}+c_1(I)]$. Since $c_1(XX)$ is also exactly known\cite{jim_korepin}, in this way we have obtained the
exact value of $c_1(I)$, which has not been known in the literature.

Our next remark concerns the entropy profile of a finite and open XX-chain, i.e. with $J_L=0$,
which has been recently calculated\cite{boundary} and a staggered behavior is
observed depending on the parity of $\ell$. According to our result
for even $\ell$ the mapping
with the QIC is perfect, whereas for odd $\ell$ there are finite-size corrections, which
scale with $1/\ell$. This result is in complete agreement with the observed
numerical findings.

For disordered chains the relation in Eq.(\ref{S_rel}) is valid for each (set of) samples,
consequently the average entropies are also related by a factor of two. If the disorder is in
the XX-form, i.e. $J_{i}^x=J_{i}^y=J_{i}$, then the effective
central charges satisfy: $c_{\rm eff}(XX)=2 c_{\rm eff}(I)$, which is also
in agreement with the results obtained by the strong disorder renormalization group method\cite{refael}.
We obtain, however, new results, if the disorder is XY-type, i.e. generally $J_{i}^x \ne J_{i}^y$.
According to our mapping
the effective central charge of random critical XY-chains is the same as for random XX-chains and
given by: $c_{\rm eff}(XY)=\ln 2$.

The effect of another type of inhomogeneities on the entanglement entropy have also
been considered. In this respect we mention numerical studies of 
the XX-chain having one or two defects, connecting the two subsystems\cite{XX_defect},
or quasi-periodic, or more generally aperiodic 
modulation of the couplings\cite{ijz07}. These perturbations result in varying effective central
charges in both models which are however related through Eq.(\ref{S_rel}).

Note added in proof: After submitting this work a paper by Cardy {\it et al} has appeared\cite{cardy}
in which $c_1(I)$ is calculated by different methods.

\acknowledgements

This work has been
supported by the National Office of Research and Technology under
Grant No. ASEP1111, by a German-Hungarian exchange program (DAAD-M\"OB), by the
Hungarian National Research Fund under Grant No OTKA TO43159,
TO48721, K62588, MO45596 and M36803.
Useful discussions with Y.-C. Lin, H. Rieger and Z. Zimbor\'as are gratefully acknowledged.


\section{Appendix: Mapping of the XY chain to two decoupled QIC-s}

This type of mapping in the thermodynamic limit is presented in\cite{peschel_schotte}
and applied for random chains in\cite{fisherxx}. For finite chains it is described in\cite{ijr00}.

Let us define two sets of Pauli operators, $\sigma_i^{x,z}$ and $\tau_i^{x,z}$, $i=1,2,\dots,L/2$
through the spin-1/2 operators $S_j^{x,y}$, with $j=1,2,\dots,L$ by:
\beqn
\sigma_i^x=\prod_{j=1}^{2i-1} (2S_j^x),\quad \sigma_i^z=4 S_{2i-1}^y S_{2i}^y \cr
\tau_i^x=\prod_{j=1}^{2i-1} (2S_j^y),\quad \tau_i^z=4 S_{2i-1}^x S_{2i}^x\;.
\label{mapping}
\eeqn
The inverse transformations are given by:
\beqn
2S_{2i-1}^x=\sigma_i^x \prod_{j=1}^{i-1} \tau_j^z,\quad 2S_{2i}^x=\sigma_i^x \prod_{j=1}^{i} \tau_j^z \cr
2S_{2i-1}^y=\tau_i^x \prod_{j=1}^{i-1} \sigma_j^z,\quad 2S_{2i}^y=\tau_i^x \prod_{j=1}^{i} \sigma_j^z\;.
\label{mapping_inv}
\eeqn
In terms of these Pauli operators the Hamiltonian operator of the XY-chain with $L$ spins in
Eq.(\ref{eq:H_XY}) can be written as the sum of two decoupled quantum
Ising chains with $L/2$ sites:
\beqn
{\cal H}_{XY}&=&\frac{1}{2}[{\cal H}_I(\sigma)+{\cal H}_I(\tau)]\cr
{\cal H}_I(\sigma)&=&-\frac{1}{2}\sum_{i=1}^{L/2} J_{2i}^x\sigma_i^x \sigma_{i+1}^x-
\frac{1}{2}\sum_{i=1}^{L/2} J_{2i-1}^y \sigma_i^z \cr
{\cal H}_I(\tau)&=&-\frac{1}{2}\sum_{i=1}^{L/2} J_{2i}^y\tau_i^x \tau_{i+1}^x-
\frac{1}{2}\sum_{i=1}^{L/2} J_{2i-1}^x \tau_i^z\;.
\label{mapping_H}
\eeqn
Here, in the last step, we have made a gauge transformation to change
the sign of the right-hand side of the last two equations. Note that although
$[{\cal H}_I(\sigma),{\cal H}_I(\tau)]=0$, the two chains are not independent
since e.g. $\sigma_i^x$ and $\tau_i^x$ do not commute with each other.



\begin{thebibliography}{0}
\bibitem{note1}
In earlier studies both in Ref.\cite{it96} and in Ref.\cite{ijr00} open chains are
considered, here we treat also closed chains.
\bibitem{note}
The matrix $\mathbf T$ has $2L$ eigenvalues, however with the transformation $\phi_k(i) \to - \phi_k(i)$
we obtain $\Lambda \to -\Lambda_k$, thus we restrict ourselves to $\Lambda_k \ge 0$, $k=1,2,\dots,L$.
\end{thebibliography}
\end{document}